\documentclass[journal=jacsat,manuscript=article]{achemso}
\setkeys{acs}{articletitle = true}
\usepackage[version=3]{mhchem} 
\usepackage[T1]{fontenc} 
\usepackage{amsmath}
\usepackage{graphicx}
\usepackage{xcolor}
\usepackage{dcolumn}
\usepackage{bm}
\usepackage{titlesec}
\usepackage{epstopdf}

\titleformat{\section}{\raggedright\fontsize{13}{25}\bfseries}{\arabic{section}.}{1em}{}

\DeclareUnicodeCharacter{0301}{'a}
\DeclareUnicodeCharacter{00D6}{"o}
\DeclareUnicodeCharacter{00C4}{"a}
\DeclareUnicodeCharacter{00FC}{"u}
\DeclareUnicodeCharacter{00DF}{\ss}
\usepackage{hyperref}

\author{Julian Wagner}
\affiliation[University of Cologne]{II. Physikalisches Institut, Universit\"at zu K\"oln, Z\"ulpicher Stra\ss e 77, K\"oln D-50937, Germany}

\author{Robin Bernhardt}
\affiliation[University of Cologne]{II. Physikalisches Institut, Universit\"at zu K\"oln, Z\"ulpicher Stra\ss e 77, K\"oln D-50937, Germany}

\author{Lukas Rieland}
\affiliation[University of Cologne]{II. Physikalisches Institut, Universit\"at zu K\"oln, Z\"ulpicher Stra\ss e 77, K\"oln D-50937, Germany}

\author{Omar Abdul-Aziz}
\affiliation[University of Cologne]{II. Physikalisches Institut, Universit\"at zu K\"oln, Z\"ulpicher Stra\ss e 77, K\"oln D-50937, Germany}

\author{ Qiuyang Li}
\affiliation[Columbia University]{Department of Chemistry, Columbia University, New York, New York 10027, United States}

\author{Xiaoyang Zhu}
\affiliation[Columbia University]{Department of Chemistry, Columbia University, New York, New York 10027, United States}

\author{Stefano Dal Conte}
\affiliation[Politecnico di Milano]{Dipartimento di Fisica, Politecnico di Milano, Piazza L. da Vinci 32, 20133 Milan, Italy}

\author{Giulio Cerullo}
\affiliation[Politecnico di Milano]{Dipartimento di Fisica, Politecnico di Milano, Piazza L. da Vinci 32, 20133 Milan, Italy}

\alsoaffiliation[CNR]{Istituto di Fotonica e Nanotecnologie, Consiglio Nazionale delle Ricerche, Piazza L. da Vinci 32, 20133 Milano, Italy
}

\author{Paul H. M. van Loosdrecht}
\affiliation[University of Cologne]{II. Physikalisches Institut, Universit\"at zu K\"oln, Z\"ulpicher Stra\ss e 77, K\"oln D-50937, Germany}
\email{pvl@ph2.uni-koeln.de}

\author{Hamoon Hedayat}
\affiliation[University of Cologne]{II. Physikalisches Institut, Universit\"at zu K\"oln, Z\"ulpicher Stra\ss e 77, K\"oln D-50937, Germany}
\email{hedayat@ph2.uni-koeln.de}

\title{Unveiling Ultrafast Spin-Valley Dynamics and Phonon-Mediated Charge Transfer in MoSe$_{2}$/WSe$_{2}$ Heterostructures}

\makeatother
\begin{document}
\begin{abstract}

We use helicity-resolved ultrafast transient absorption spectroscopy to study spin-valley polarization dynamics in a vertically stacked MoSe$_{2}$/WSe$_{2}$ heterostructure. The experimental findings reveal details of interlayer charge transfer on ultrafast timescales, showing that the spin-valley polarized state of photoexcited carriers is conserved during the charge transfer and formation of interlayer excitons. Our results confirm that phonon scattering mediates the interlayer charge transfer process, while a high phonon population at elevated temperatures causes a significant decrease in spin-valley selective charge transfer. Moreover, the experimental findings demonstrate the possibility that interlayer excitons and their spin-valley polarization can be probed in the optical response of intralayer excitons. These findings pave the way for ultrafast detection, control, and manipulation of spin-valley polarized excitons in transition metal dichalcogenide-based 2D heterostructures. 

\end{abstract}

\maketitle

\section[INTRODUCTION]{INTRODUCTION\label{Introduction}}

Semiconducting monolayers of transition metal dichalcogenides (TMDs) show a variety of interesting optical and excitonic properties~\cite{splendiani2010emerging,jin2018ultrafast,xiao2017excitons}. These materials exhibit a direct optical gap, large exciton binding energies (i.e. 100s meV), and valley-selective optical properties originating from spin-orbit coupling combined with a broken inversion symmetry in the monolayer limit~\cite{splendiani2010emerging,jin2018ultrafast,xiao2017excitons}. However, a drawback for potential applications is that excitons and spin-valley polarization in TMD monolayers have relatively short lifetimes~\cite{splendiani2010emerging,jin2018ultrafast,xiao2017excitons}. A possible route to increase these lifetimes is to make use of stacked monolayer systems. The weak van der Waals forces between the layers allow vertical stacking of two TMD monolayers to form vertical heterostructures (HSs)~\cite{novoselov20162d} in which the electronic structure of the individual layers is preserved close to the \textit{K} valleys. Of particular interest are HSs exhibiting a type II band alignment, as found in MoSe$_{2}$/WSe$_{2}$, in which the valence band maximum and the conduction band minimum reside in different layers~\cite{liu2020excitons,novoselov20162d}. This gives rise to ultrafast sub-picosecond (sub-ps) and efficient interlayer charge transfer (CT) after optical excitation of one of the layers, with holes accumulating in the valence band of WSe$_{2}$\ and electrons in the conduction band of MoSe$_{2}$~\cite{hong2014ultrafast}. The sub-ps CT observed in TMD HSs is considered to be caused by a combination interlayer-hybridization, large density of states for resonant interlayer CT with type II band alignment, and phonon-mediated scattering~\cite{wang2021phonon,wang2016role,zimmermann2020directional, chen2016ultrafast,liu2020direct,li2017charge}. While the spatial separation of electrons and holes in the two distinct layers suppresses electron-hole recombination due to reduced electron-hole wavefunction overlap~\cite{rivera2016valley}, the strong electron-hole Coulomb attraction gives rise to the formation of interlayer excitons (ILEs) with binding energies of a few 100 meV and emission energies lower than those of intralayer excitons~\cite{jiang2021interlayer}. 
The impact of the relative twist angle between the constituent monolayers on the ultrafast CT process has been intensively studied~\cite{volmer2023twist,jiang2021interlayer,DALCONTE202028,doi:10.1021/acs.nanolett.7b00748, doi:10.1021/acsnano.7b04541,zimmermann2021ultrafast,li2022spin}. 
Despite these efforts, there are still important open questions on interlayer CT process and subsequent formation process of ILEs, in particular whether spin-valley polarization from the initially excited intralayer excitons is transferred to free charge carriers. Further, the impact of spin-valley polarization on the CT process is yet to be explored.

In contrast to the ultrafast CT process, the recombination dynamics, as expected, is much slower, with reported timescales ranging from nanoseconds to microseconds for single carrier spin lifetime in doped and gated TMD monolayer~\cite{ersfeld2020unveiling,yang2015long} and the ILE lifetime in TMD HSs~\cite{jiang2021interlayer,DALCONTE202028}, influenced by factors such as the temperature and the Moir\'e order~\cite{schaibley2016directional}. Importantly, increased spin-valley lifetimes have also been reported in TMD HSs \cite{kim2017observation}, which is promising for valleytronics applications~\cite{rasmita2021opto,liu2019valleytronics}. 

Here we investigate the dynamics and scattering processes of valley-polarized carriers in a MoSe$_{2}$/WSe$_{2}$\ HS using helicity-resolved broadband transient absorption (TA) spectroscopy. We find that CT is driven by phonon scattering at elevated temperatures, resulting in a significant loss of spin-valley polarization. Conversely, at low temperatures, where phonon scattering events are limited, carriers maintain their spin-valley polarization and form valley-polarized ILEs as they transfer between layers. Although it is difficult to optically detect ILEs due to their low oscillator strength, our investigation indicates the possibility of probing ILEs and their spin-valley polarization by analyzing much stronger transient optical responses of the intralayer excitons.

\section[RESULTS AND DISCUSSION]{RESULTS AND DISCUSSION}\label{results}

\begin{figure}[ht]
\includegraphics[width=\textwidth]{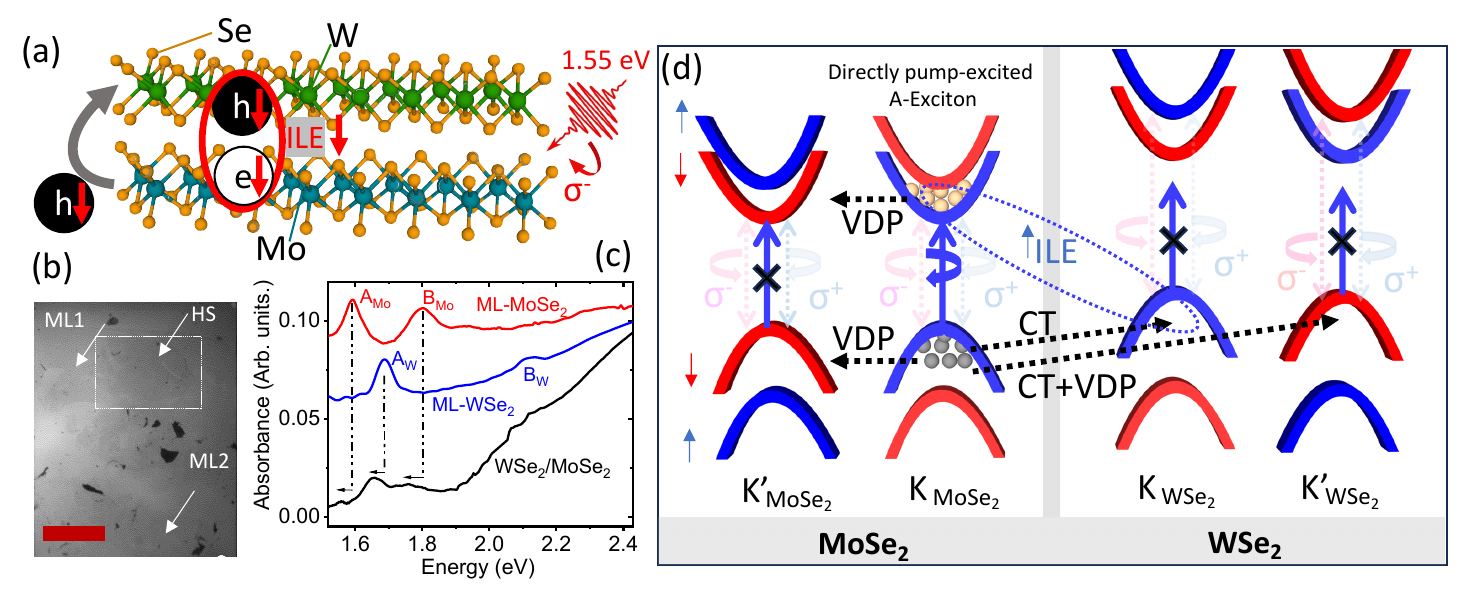}

\caption{ (a) Schematics of the TMD HS composed of MoSe$_{2}$\ and WSe$_{2}$\ monolayers, where selective excitation promotes valley-polarized hole transfer.
(b) Optical image of the MoSe$_{2}$/WSe$_{2}$\ HS sample, with arrows indicating the location (scale: 5 mm). White dashed box shows the HS area. (c) Static absorption confirms the presence of the HS in the measured spot at room temperature, with observable shifts in exciton energy positions compared to the monolayers. For comparison purposes, the spectra are shifted along the y-axis.
(d) Sketches of the valence and conduction band type II alignment at the K and K' valleys in the HS. The blue and red colors denote contrasting valley states. The dark blue arrows represent the 1.55 eV pump that is a right circularly polarized pulse at resonant with the A-exciton of MoSe$_{2}$, selectively promoting electrons in a \textit{K} valley of MoSe$_{2}$. The light arrows show the two opposite circularly polarized probes which reveal the dynamics of valley-polarized carriers in each valley. Black dashed arrows indicate the transfer of charge (CT), while VDP represents the valley depolarization process, both processes lead to the indirect population of different valleys at longer time delays.}
\label{Fig1}
\end{figure}

We study a large area MoSe$_2$/WSe$_2$ HS in which the monolayers have been obtained by gold-film-assisted mechanical exfoliation of bulk crystals~\cite{liu2020disassembling}. Using angle-resolved second harmonic-generation microscopy, the twist angle between the monolayers is determined to be $\approx 2^{\circ}$ (see Supplementary Information (SI) for more details). The stacking order of the individual monolayers and the possible valley-selective hole transfer after resonantly driving/pumping a specific valley in MoSe$_{2}$ with circularly polarized light are shown in Fig.~\ref{Fig1}(a). 
Fig.~\ref{Fig1}(b) shows an optical image of the sample that displays the mm-sized monolayers of MoSe$_{2}$\ and WSe$_{2}$\ and the HS area that is visually discernible to the naked eye. Raman spectroscopy measurements have been conducted to prove the monolayer nature of the distinct layers (see SI).
Further, Fig.~\ref{Fig1}(c) reports the static optical absorption spectrum of the HS and the isolated TMD monolayers at room temperature. 
The absorption spectra for both individual monolayers and HS samples reveal the expected spectral signatures of the intralayer excitonic transitions labeled as A$_X$, B$_X$\ with $X$=Mo,W~\cite{hsu2018negative}. The linear steady state absorption spectrum of the HS shown in Fig.~\ref{Fig1}(c) appears to be approximately the superposition of the spectra for the individual monolayers. However, all excitonic transitions are spectrally broadened, which has been assigned to the additional decay channel of the intralayer exciton to the CT processes between the layers. This indicates a good coupling of the constituent single monolayers \cite{rigosi2015probing,wang2016interlayer}. Furthermore, excitonic resonances are slightly redshifted due to the different dielectric environment of the TMDs in the HS \cite{schaibley2016directional}. The A-exciton feature corresponds to the energetically lowest band-edge transition at the \textit{K} and \textit{K'} points of the hexagonal Brillouin zone~\cite{glazov2021valley}. Fig.~\ref{Fig1}(d) illustrates the schematic type II band alignment of the heterobilayer and the corresponding spin-valley polarization of each band, where the energetically lowest (highest) conduction (valence) band is located in the MoSe$_{2}$ (WSe$_{2}$) monolayer, respectively. 
We perform helicity resolved broadband TA measurements with right-circularly ($\sigma^+$) polarized pump pulses tuned to the A-exciton resonance of MoSe$_{2}$\ (1.55 eV) as shown in Fig.~\ref{Fig1}(d), while the probe pulse is either right- ($\sigma^+ \sigma^+$) or left- ($\sigma^+ \sigma^-$) circularly polarized light (see SI for additional pump photon energies). We probe the valley contrast through the circular dichroism (CD) defined as $\text{CD}(t)=\Delta A_{\sigma^{+}\sigma^{+}}(t)-\Delta A_{\sigma^{+}\sigma^{-}}(t)$. Here, $\Delta A_{\sigma^{+}\sigma^{+}}(t)$ is the TA response for the same pump and probe helicities and $\Delta A_{\sigma^{+}\sigma^{-}}(t)$ for opposite helicities~\cite{schaibley2016directional}. 

Figure.~\ref{Fig1}(d) illustrates all the scattering processes occurring in the HS after resonant photoexcitation at the \textit{K} valley of MoSe$_2$ with circularly polarized light. Due to the valley-dependent optical selection rules, the photoexcitation with pump pulses resonant with the optical gap of MoSe$_2$ results in the creation of spin-polarized electron and hole populations in the \textit{K} valley of MoSe$_2$. The type II band alignment of the HS favors the scattering of spin-polarized holes from MoSe$_2$ to WSe$_2$.~\cite{rivera2015observation,zimmermann2021ultrafast}. After the charge transfer process, the recombination timescale of spatially separated electrons and holes located in different layers is orders of magnitude longer than that of carriers located in the same layer. This is due to the reduced spatial overlap between the electron and hole wavefunctions. Since CT breaks the intralayer excitons, it is expected also to reduce the electron-hole exchange interaction, which is responsible for the fast (i.e., sub-ps) valley depolarization dynamics observed in single-layer TMDs. For this reason, the spin/valley relaxation time of spin-polarized electrons in the conduction band (CB) of MoSe$_2$ and holes in the valence band (VB) of WSe$_2$ is expected to increase with respect to that measured in isolated layers. 

\begin{figure*}[ht]
\includegraphics[width=\textwidth]{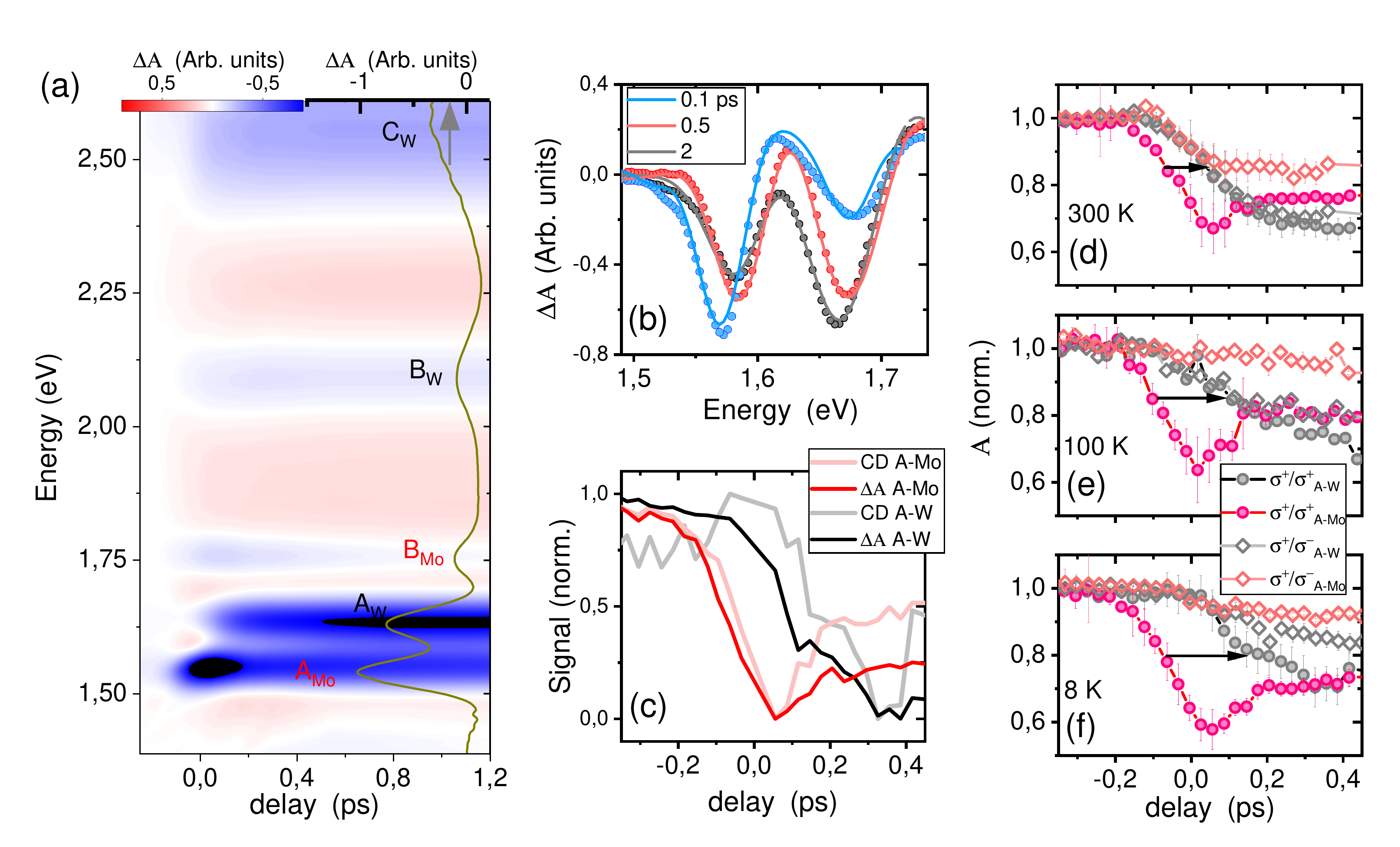}

\caption{ (a) 2D map of the transient absorption obtained for excitation in resonance with $A_{\text{Mo}}$. The black curve shows all excitonic features at 0.2 ps delay.  (b) shows the transient absorption data at different time delays of 0.1 ps, 0.5 ps, and 1 ps, along with their corresponding fits. Using the fit results one can extract the absorption amplitude of $A_{\text{Mo}}$\ and $A_{\text{W}}$\ at each delay (A(t)). At early delay times, $A_{\text{Mo}}$\ dominates the TA signal, while at longer times, $A_{\text{W}}$\ becomes more pronounced due to CT between layers. (c) shows the dynamics of $A_{\text{Mo}}$\ and $A_{\text{W}}$\ excitons (dark colors) alongside the corresponding CD dynamics (light colors) at 8K. (d), (e), and (f) depict the normalized absorption amplitude A(t) dynamics, extracted by the fitting procedure shown in (b), at temperatures of 300 K, 100 K, and 8 K, respectively. At all three temperatures, $A_{\text{Mo}}$\ of the \textit{K} valley, excited by $\sigma^+$/$\sigma^+$\ polarization, initially responds, followed by the response of other excitons in the \textit{K} and \textit{K'} valleys of each layer with some delay. By lowering the temperature, the transfer exhibits a distinct delay as shown by arrows and discussed in the text.
}
\label{Fig4}
\end{figure*}

Fig.~\ref{Fig4}(a) shows the sub-ps TA spectra obtained after photoexcitation of the MoSe$_{2}$/WSe$_{2}$\ HS using circularly polarized pump and probe ($\sigma^+\sigma^+$) in resonance with A$_{\text{Mo}}$. The chirp in the TA spectra has been corrected using a reference measurement of the substrate only (see SI). The black curve illustrates the TA spectrum at 0.2 ps delay time, highlighting the distinct A, B and C excitonic resonances of MoSe$_{2}$ and WSe$_{2}$\ in the HS. The 2D color map is dominated by the bleaching of the intralayer exciton peaks, i.e., A$_{\text{Mo}}$ and A$_{\text{W}}$. While A$_{\text{Mo}}$ rises instantaneously, A$_{\text{W}}$ is delayed in time.  We focus on the A-excitons and present three spectra at early time delays of 0.1, 0.5, and 2 ps together with their respective fits in Fig.~\ref{Fig4}(b) (the fitting procedure is described in SI). We use the fitm results to more effectively disentangle the dynamics of A$_{\text{Mo}}$\ and A$_{\text{Mo}}$\ and to extract the amplitude of absorption (A(t)) of both intralayer excitons at each time delay. An initial enhanced bleaching signal of the A$_{\text{Mo}}$\ feature at 0.1 ps delay and a subsequent transfer of the spectral weight to A$_{\text{W}}$\ on the sub-ps timescale is observed. A$_{\text{Mo}}$\ rises instantaneously, while A$_{\text{W}}$\ is delayed due to  interlayer hole transfer. Fig.~\ref{Fig4}(c) shows that the bleaching of the A$_{\text{W}}$\ transition (black line) is delayed after the A$_{\text{Mo}}$\ is bleached (dark red line). In addition, the CD of each excitonic transition is shown in a lighter color and follows the bleaching dynamics. This indicates that the transient spin-valley polarization and the hot carrier dynamics evolve closely at low temperatures. 

The dynamics of the A-exciton is studied as a function of the temperature in order to investigate the possible role of phonon mediated scattering and spin valley CT processes (cf. Fig.~\ref{Fig4}(d), \ref{Fig4}(e), and \ref{Fig4}(f)). The choice of these three temperatures (8, 100 and 300 K) was made to aid in the investigation of the role of phonons, given that phonon activation occurs around 100~K (see SI for detailed temperature-dependent measurements), consistent with previous reports~\cite{yan2017exciton,miyauchi2018evidence}. The CT to the WSe$_{2}$\ layer is confirmed by the delayed rise of the A$_{\text{W}}$\ exciton features. The risetime of A$_{\text{W}}$ excitons aligns well with the theoretically predicted CT process, which is estimated to occur with a time constant of 100s fs, which is started with an ultrafast interlayer hopping process in a few tens of fs \cite{zeng2021new}. There is an increase in the delay of the rise time of A$_{\text{W}}$ as the temperature decreases (black arrows indicate a delay of about 50 fs at 300 K, which extends to about 200 fs at 8 K). A similar delay in CT dynamics was detected in a comparable TMD HS using time-resolved Kerr spectroscopy~ \cite{kumar2021spin}. The finite rise time is attributed to the phonon-assisted CT process, where optical phonons with energy in the tens of meV range facilitate the interlayer scattering of holes~\cite{miyauchi2018evidence,yan2017exciton,mahrouche2022phonons}. The reduced efficiency of phonon-assisted CT at 8 K is attributed to the decrease in thermally active phonons (see SI) impacting the adiabatic CT mechanism~\cite{selig2019ultrafast,zheng2017phonon}, which explains the observed delay in the rise of the A$_{\text{W}}$ signal as temperature decreases. Therefore, our results confirm the phonon-mediated CT mechanism discussed in other studies~\cite{policht2023timedomain,wang2021phonon,wang2016role}.  More interestingly, it is shown that the dynamics of A$_{\text{W}}$ is valley-independent when the CT is mediated by high phonon scatterings, as evident in Fig.~\ref{Fig4}(d and e) where A$_{\text{W}}$ at \textit{K} (filled black circles) and A$_{\text{W}}$ at \textit{K'} (hollow black squares) follow the same dynamics. At 8~K (Fig.~\ref{Fig4}(f)), in contrast, the rise times of the bleaching of A$_{\text{W}}$ at \textit{K} and \textit{K'} are different, indicating that CT is not identical for the two valleys, suggesting spin-valley polarized CT. While phonons mediate both CT~\cite{zheng2017phonon} and valley depolarization (VDP)~\cite{lin2022phonon} in TMD HSs, our results indicate that CT remains more effective at low temperatures, due to interlayer coupling involving electronic and valley states. This coupling conserves valley polarized CT efficiency, in contrast to VDP, which relies more strongly on phonon-assisted spin-flip processes. To further clarify this point, we analyze the rise time of the A$_{\text{Mo}}$\ and A$_{\text{W}}$\ transition feature probed in co- and cross-polarized configurations in order to evaluate the VDP and CT rates. We aim to determine how the transferring holes preserve their spin-valley polarization during interlayer CT. 

To gain deeper insights into these complex processes, we apply a rate equation based model. The model takes into account the four different exciton states residing in the different monolayers and \textit{K}/\textit{K'} valleys;  N$_{Mo^+}$\ at \textit{K}$_{\text{MoSe}_{2}}$, N$_{W^+}$\ at \textit{K}$_{\text{WSe}_{2}}$, N$_{W^-}$\ at \textit{K'}$_{\text{WSe}_{2}}$, and N$_{Mo^-}$\ at \textit{K'}$_{\text{MoSe}_{2}}$. As discussed earlier and since our pump photon energy is well below the A exciton transition of WSe$_2$, we assume that the ultrafast bleaching features of the WSe$_2$ transitions mainly originate from hole transfer from the MoSe$_2$ to the WSe$_2$ layer. Additionally, we assume that only the sub-ps dynamics of the directly photoexcited A$_{\text{Mo}}$\ exciton at \textit{K} serves as the source of both inter- and intra-layer spin-valley charge dynamics. Since the CT process, occurring in a few hundred femtoseconds, is significantly faster than radiative recombination, we exclude the latter from the rate equations \cite{may2020nanocavity}. To simplify the model, we choose not to include separate equations for electron dynamics in the \(\text{MoSe}_2\) layer, assuming instead that electrons influence the A$_{\text{Mo}}$ absorption feature through phase space filling. In the following, we demonstrate that this simplified model can reproduce the ultrafast dynamics and extract crucial information on spin-polarized hole transfer.

\begin{figure*}[ht]
\includegraphics[width=0.9\textwidth]{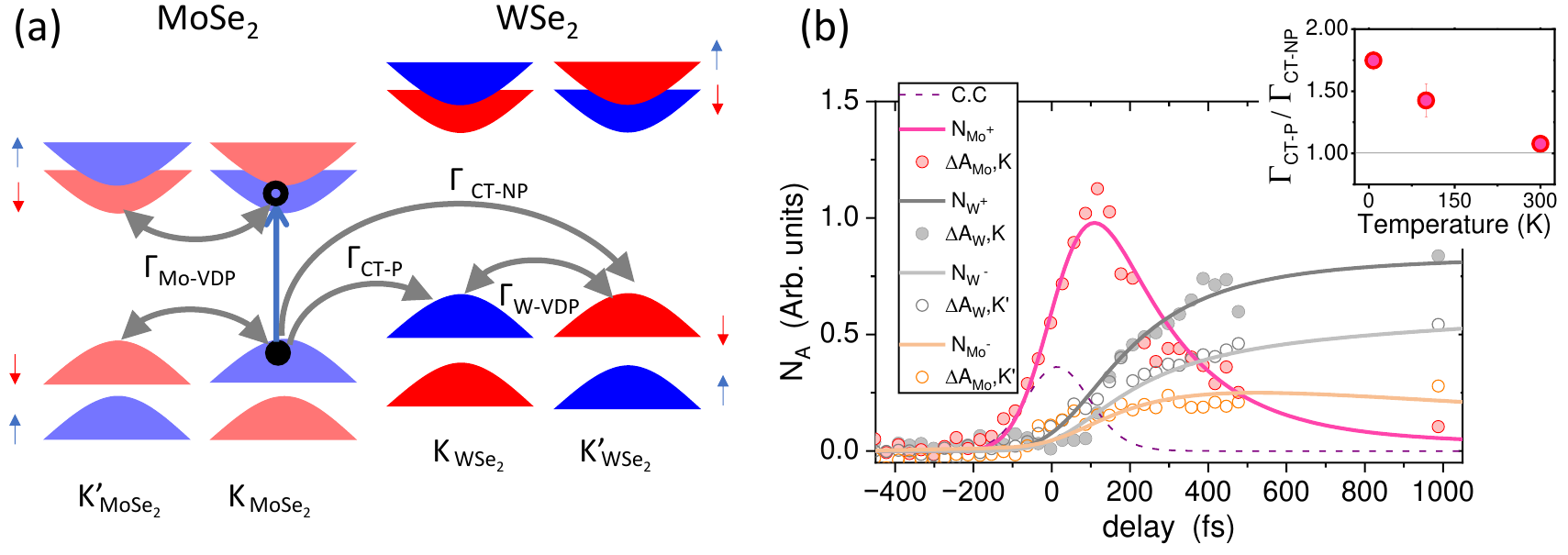}
\caption{ (a) The sketch illustrates the carrier dynamics and potential decay channels for excited valley-polarized holes. The holes can be transferred to the other layer while maintaining their polarization ($\Gamma_\text{CT,P}$), or transfer and subsequently depolarize in the other layer ($\Gamma_\text{CT,NP}$). Additionally, VDP can also occur within the same layer ($\Gamma_\text{Mo,VDP}$) and ($\Gamma_\text{W,VDP}$). By employing the rate equations discussed in the main text, the experimental TA data at 8~K can be fitted as shown in panel (b). The results highlight the crucial role of spin-valley polarization in the CT process. The inset shows the obtained ratio of $\Gamma_\text{CT,P}/\Gamma_\text{CT,NP}$ at different temperatures.}

\label{Fig5}
\end{figure*}

In the rate equations, the excitation in resonance with the A-exciton transition in MoSe$_2$\ at \textit{K}, responsible for the creation of sub-ps transferring carriers, is defined by a Gaussian source term, i.e. $N_{0}(t)$, which is determined by the instrumental response function, i.e. the cross-correlation of the pump and probe pulses, with a FWHM of 100 fs. We directly include it in the first equation as $N_{0}(t)$, which serves as the source of all other dynamics. Subsequently, we investigate the dynamics of all \textit{K} and \textit{K'} valleys and derive the scattering rates between them. When focusing solely on the MoSe$_2$\ layer, we can define a depolarization channel from the excited valley \textit{K} to the \textit{K'} valley, referred to as VDP rate of the MoSe$_2$\ layer ($\Gamma_\text{VDP,Mo}$). Similarly, we can define an analogous depolarization channel in the WSe$_2$\ layer ($\Gamma_\text{VDP,W}$). Moreover, we consider that the hole transfer between layers can potentially occur through valley-preserving ($\Gamma_\text{CT,P}$) or non-valley-preserving channels ($\Gamma_\text{CT,NP}$). Fig.~\ref{Fig5}(a) schematically illustrates the coupling between the valleys and the scattering channels. The dynamics of the system is described by the following rate equations,
\begin{align*}
\frac{dN_{Mo^+}}{dt} &= N_0 - \Gamma_\text{CT,P}N_{Mo^+} - \Gamma_\text{CT,NP}N_{Mo^+} + \Gamma_\text{VDP,Mo}(N_{Mo^-} - N_{Mo^+}) \\
\frac{dN_{W^+}}{dt} &= \Gamma_\text{CT,P}N_{Mo^+} + \Gamma_\text{VDP,W}(N_{W^-}-N_{W^+}) \\
\frac{dN_{W^-}}{dt} &= \Gamma_\text{CT,NP}N_{Mo^+} + \Gamma_\text{VDP,W}(N_{W^+}-N_{W^-}) \\
\frac{dN_{Mo^-}}{dt} &= \Gamma_\text{VDP,Mo}(N_{Mo^+} - N_{Mo^-}).
\end{align*}
The results, as depicted in Fig.~\ref{Fig5}(b), reproduce the experimentally observed dynamics at 8~K reasonably well. In the following, we discuss the derived parameters related to the CT processes in detail. The findings reveal that the valley preserving CT rate in WSe$_{2}$ ($\Gamma_\text{CT,P} \approx 2.08\pm0.02~\text{ps}^{-1}$) is higher than the rate considering a transfer to the \textit{K'} valley ($\Gamma_\text{CT,NP} \approx 1.19\pm0.05~\text{ps}^{-1}$). As a result, after the CT process is completed, we observe the same valley polarization in both layers, i.e., the same CD sign at long delays, in agreement with previous studies \cite{hsu2018negative}. This can be rationalized by the fact that scattering to an oppositely polarized valley would require a spin flip, and further is impeded by the large spin-orbit splitting of the valence bands. The observed higher CT rate of valley-polarized charge carriers compared to the CT rate of non-valley-polarized carriers underscores the importance of valley degrees of freedom in the CT dynamics. Furthermore, the analysis shows that the CT rate from MoSe$_{2}$\ to the opposite valley in WSe$_{2}$\ ($\Gamma_\text{CT,NP} \approx 1.19\pm0.05~\text{ps}^{-1}$) is higher than the depolarization rate in the same layer ($\Gamma_\text{VDP,Mo} \approx 0.83\pm0.17~\text{ps}^{-1}$), indicating that CT is the dominant process on the sub-ps timescale rather than intralayer VDP. We use the rate equations for examining the dynamics under different temperature conditions (refer to SI). The inset of Fig.~\ref{Fig5}(b) shows that the ratio of $\Gamma_\text{CT,P}/\Gamma_\text{CT,NP}$ decreases at higher temperatures, implying that at elevated temperatures, spin-polarized holes in WSe$_2$ can scatter to the \textit{K} or \textit{K'} valleys with equal probability. Consequently, the phonon-mediated CT process washes out the spin-valley polarization of the CT.

\begin{figure*}[ht]
\includegraphics[width=0.85\textwidth]{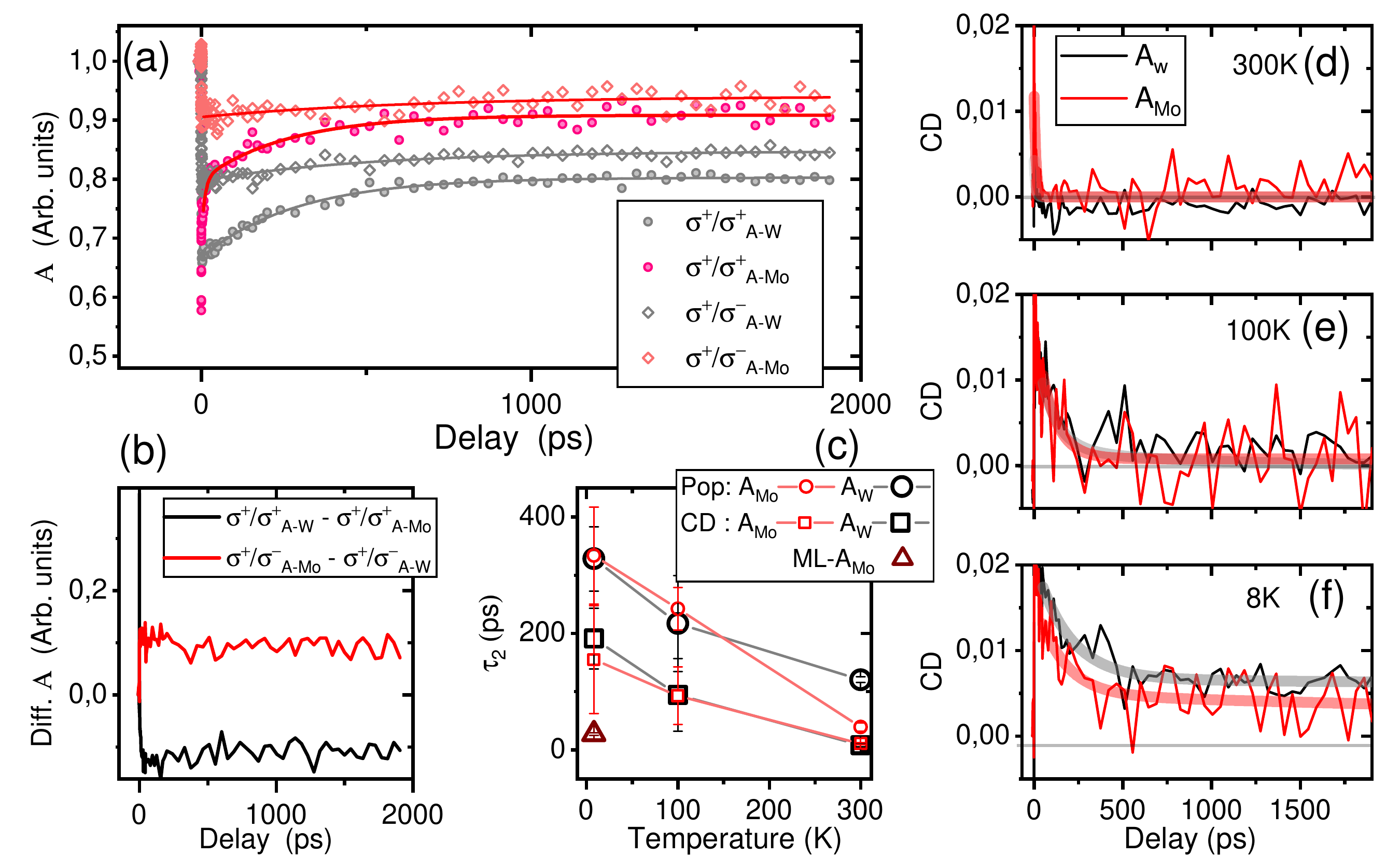}
\caption{ (a) At long delays in the nanosecond range, the dynamics of the transient absorption amplitude of $A_{\text{Mo}}$\ and $A_{\text{W}}$\ are similar. The solid lines indicate the exponential fitting showing similar decay rates. (b) shows the difference between $A_{\text{Mo}}$\ and $A_{\text{W}}$\ of the same \textit{K} or \textit{K'} valleys, which remain constant in the ns time domain range. (c) exhibits the decay time constants for the population (circles) and CD (squares) of $A_{\text{Mo}}$ (in red) and $A_{\text{W}}$ (in black). The brown triangle shows the population decay time constant of  $A_{\text{Mo}}$ monolayer. (d)-(f) The transient CD signal of MoSe$_{2}$\ and WSe$_{2}$\ show the same dynamics at different temperatures. The pale-colored curves represent the exponential fitting applied to the CD data.}

\label{Fig6}
\end{figure*}

Finally, we discuss the dynamics on longer timescales, where the exciton population decays through recombination (see Fig.~\ref{Fig6}(a)). The photobleaching signals of A$_{\text{W}}$ and A$_{\text{Mo}}$ show comparable decay dynamics as shown in Fig.~\ref{Fig6}(b). After the initial subpicosecond CT dynamics, the decay dynamics is fitted using a triexponential function. The first time constant ($\tau_{1}$$<30$~ps) is present only for A$_{\text{Mo}}$ at \textit{K} valley. We attribute this to the recombination of intralayer excitons that were initially created and not transferred, which then recombine within the MoSe$_2$ layer over a few tens of picoseconds. The recombination mechanisms for intralayer excitons include both radiative and non-radiative processes. Among the radiative mechanisms are scattering with defects and exciton-exciton annihilation, as supported by earlier studies \cite{kumar2014exciton}. The most extended time scale ($\tau_{3} >1.8$~ns) at 8~K surpasses the maximum delay interval that can be accessed in our experiments, hence, we are unable to assign an exact value to it, given that it can range up to tens of ns and potentially more. The second time constants ($\tau_{2}$) are plotted in Fig.~\ref{Fig6}(c). We observe a very similar decay time for A$_{\text{Mo}}$ and A$_{\text{W}}$. The observed dynamics may originate from the following three processes: (i) the recombination of the intralayer A-exciton in individual monolayers, as discussed for $\tau_{1}$ of A$_{\text{Mo}}$, through mechanisms such as Auger recombination \cite{wang2015ultrafast}, (ii) bimolecular recombination of electrons and holes residing in different layers, i.e., after CT, electrons and holes occupy the valence and conduction band states in each layer and are spatially separated, but they have not formed ILEs and recombine or diffuse independently.
(iii) the geminate recombination of bound electrons and holes residing in separate monolayers (i.e., ILE recombination). All these processes in principle contribute to the decay of the intralayer excitons TA signal due to phase space filling mechanisms. In the following, we examine the possible origin of the decay rates observed in the A-excitonic resonances.

The first possibility can be easily ruled out since the analysis performed on the sub-ps timescale demonstrates that the dominant hole transfer populates the layers with different hot charge carriers (electrons and holes), each with their own decay dynamics. Nevertheless, the decay behavior remains consistent for the corresponding valleys of \textit{K} (or \textit{K'}) in different layers as the difference in their decay is fairly constant (see Fig.~\ref{Fig6}(b)). Figure.~\ref{Fig6}(c)  shows the similar decay time constants of both layers. In the context of the second scenario, the presence of intralayer long-lived defect states is ruled out. Fig.\ref{Fig6}(c) (triangle symbol) shows that A$_{\text{Mo}}$ of the monolayer does not exhibit such long-lasting localized states~\cite{ersfeld2019spin, yang2015long}, as it displays a significantly faster decay at 8~K. It can be speculated that separated electrons and holes live as resident states, i.e. remaining carriers subsequent to the recombination events ~\cite{plechinger2016trion,hsu2015optically},  and relax across the interface of the layers over extended timescales.  However, Figs.~\ref{Fig6}(d)-(f) reveal that the CDs taken from both layers display similar dynamics at different temperatures. The time constants of CD decay (spin-valley polarization) are shown in Fig.~\ref{Fig6} (c), along with the time constants for population decay. It is known that the large spin-orbit splitting in the valence band, which measures hundreds of meV, impedes spin-polarized hole scattering and makes it less probable compared to the electron VDP in the conduction bands with only about 10 meV spin-orbit splitting~\cite{kosmider2013large}. We observed that the electron-populated layer MoSe$_2$\ and the hole-populated layer WSe$_2$\ show similar spin-valley relaxation time scales in optical responses. A persistent ns spin-valley polarization is found at 8~K for both A$_{\text{Mo}}$ and A$_{\text{W}}$, in agreement with the spin-valley lifetime of the ILEs~\cite{nagler2017interlayer}.
On the other hand, the spin-valley polarization of the A$_{\text{Mo}}$ for monolayer MoSe$_2$ lasts only $20\pm2$ ps (refer to SI), which is at least two orders of magnitude less than for the HS. Thus, a common origin for the decay of the A-exciton bleaching of both layers is present, occurring when the spatially separated electrons and holes bind to form ILEs. As shown in Fig.~\ref{Fig6}(c), a notable slowdown in population decay is observed at lower temperatures, which is associated with a decrease in electron-phonon coupling, in line with the nonradiative recombination of electron-hole pairs, i.e. ILEs, as discussed in Ref.~\cite{wang2021phonon}. Thus, we reveal the potential to investigate not only the population of ILEs but also their spin-valley polarization by monitoring the much stronger transient signal of the intralayer excitons. After the initial creation of valley-polarized intralayer excitons by resonant photoexcitation followed by ultrafast CT between the layers, the long-time dynamics likely indicates the presence of bound electrons and holes in the valence and conduction bands of the two adjacent layers, which drives the formation of ILEs~\cite{rivera2015observation,hanbicki2018double}. Recent studies reported on the direct optical detection of ILEs without focusing on spin-valley polarization~\cite{barre2022optical,policht2023timedomain}. Although ILEs dominate the photoluminescence spectrum~\cite{rivera2015observation}, detecting them in the optical response is quite challenging because their oscillator strength is more than two orders of magnitude weaker than that of intralayer excitons. In this regard, the results show fingerprints of the dynamics of the ILE in the near-visible optical range with a spin-valley polarization preserved over nanoseconds after ultrafast CT.

\section[CONCLUSIONS]{CONCLUSIONS\label{Conclusions}}

In this study, the dynamics of spin-valley polarization in a MoSe$_{2}$/WSe$_{2}$\ HS have been investigated using ultrafast broadband and helicity-resolved TA spectroscopy. Resonant excitation with circularly polarized light selectively populates the \textit{K} valleys in MoSe$_{2}$ and the subsequent hot charge carrier transfer across the HS is tracked on ultrafast timescales. These results quantitatively confirm that type II TMD HSs can be used to take advantage of valley-polarized states along with ultrafast charge separation phenomena. The study focuses on details of valley-selective CT as the primary mechanism for hole transfer at low temperatures. However, at higher temperatures, the hole that scatters from one layer to the other loses spin-valley polarization. Furthermore, the results indicate that the ILEs and their spin-valley polarization play a role in the bleaching of intralayer A-excitons of the HS over longer time delays. These insights into the ultrafast spin-valley dynamics and phonon-mediated charge transfer in TMD-based HSs could inform the design of advanced optoelectronic and valleytronic devices. By understanding the conditions that preserve spin-valley polarization, these HSs may be engineered for applications such as spin-valley filters and components for quantum computing. Future studies might focus on exploring external controls to manipulate spin-valley dynamics, thereby enhancing their potential for practical applications in the development of valleytronic technologies.

\section[METHODS]{METHODS\label{experiment}}

{\bfseries{1) Sample growth, characterization and orientation}}.\\

MoSe$_{2}$\ and WSe$_{2}$\ monolayers are obtained by gold-assisted mechanical exfoliation from bulk MoSe$_{2}$\ (SPI Supplies) and WSe$_{2}$\ (HQ graphene). A gold layer is deposited on top of the bulk TMDs. When the gold layer is pulled off with a thermal release tape (Semiconductor Corp.), it carries a large piece of TMD monolayer on the contact surface. Heating up to 130°C removes the thermal release tape, and the residues are cleaned by acetone and O$_{2}$\ plasma treatment. Finally, a gold etchant solution made from mixing KI (99.9\%, Alfa Aesar) and I$_{2}$\ (99.99\%, Alfa Aesar) in deionized water is used to dissolve gold. The TMD monolayers are comparable to the monolayers obtained from conventional scotch tape exfoliation in terms of a clean surface and strong photoluminescence. MoSe$_{2}$\ is first exfoliated and transferred onto a 200 $\mu$m-thick SiO$_{2}$\ substrate. Then, WSe$_{2}$\ is exfoliated and transferred on top of MoSe$_{2}$/SiO$_{2}$.\\

{\bf{2) Experimental procedures and measurement technique}}.\\

For the ultrafast transient absorption spectroscopy measurements, a Yb:KGW regenerative amplifier laser system (PHAROS, LightConversion) 1450~nm center wavelength for Wl generation, and 75 kHz repetition rate was used. The excitation pulses at 750 nm (1.65 eV) and 800 nm (1.55 eV) were obtained through frequency conversion using a non-collinear optical parametric amplifier (TOPAS, LightConversion), while the whitelight probe pulses were generated by focusing the 1450~nm laser line on a sapphire crystal and covered a spectral range of 500 to 850 nm (1.45 to 2.5 eV). A motorized delay stage created time delays up to 1800 ps between pump and probe pulses. The transmitted whitelight probe was detected using a spectrometer with CCD detectors (Hamamatsu S8380-256Q with a readout circuit c7884). The pump and probe pulses were overlapped on the sample at a small angle, with the pump blocked afterward and the probe directed into the spectrometer. For the pump beam, a Berek compensator was used to generate circularly polarized light, while a broadband $\lambda/4$-plate was used to generate circularly polarized light for the whitelight probe.

\section*{ACKNOWLEDGMENTS\label{ack}}
The authors acknowledge financial support funded by the Deutsche Forschungsgemeinschaft (DFG) through project No. 277146847-CRC1238, Control and Dynamics of Quantum Materials (subproject No. B05). G.C. acknowledges financial support by the European Union’s NextGenerationEU Programme with the I-PHOQS Infrastructure [IR0000016, ID D2B8D520, CUP B53C22001750006] "Integrated infrastructure initiative in Photonic and Quantum Sciences". G.C. acknowledges funding from the European Horizon EIC Pathfinder Open programme under grant agreement no. 101130384 (QUONDENSATE). R.B. an L.R. acknowledge financial support by RTG-2591 "TIDE-Template-designed Organic Electronics". Sample preparation was supported by the Materials Science and Engineering Research Center (MRSEC) through NSF grant DMR-2011738 (to X.Y.Z).

\section*{Additional Information}
\begin{itemize}
\item Competing interests: The authors declare no competing interests.

\item Availability of data and materials: Data pertaining to this work is available by following link:
\item Authors' contributions: H.H., S.D.C., G.C. and P.v.L. conceived the project. J.W., R.B., and L.R. performed the transient absorption experiments. J.W. analyzed the data. H.H. and P.v.L. supervised the experiments and analysis. O.A.A. performed Raman measurements. Q.L. and X.Y.Z. provided TMD HS samples. All authors contributed to the discussion and interpretation of the results. H.H. and J.W. prepared the manuscript with input from all authors.
\end{itemize}

\clearpage

\clearpage

\clearpage

\clearpage
\def\bibsection{\section*{~\refname}} 
\bibliography{bibliography}

\end{document}